# Direct observation of the spin-orbit coupling effect in Magnetic Weyl semimetal Co$_3$Sn$_2$S$_2$

D. F. Liu[1,2*], E. K. Liu[3,4*], Q. N. Xu[5*], J. L. Shen[3], Y. W. Li[6], D. Pei[6], A. J. Liang[2,7], P. Dudin[8], T. K. Kim[8], C. Cacho[8], Y. F. Xu[1], Y. Sun[5], L. X. Yang[9,10], Z. K. Liu[2,7], C. Felser[5,11,12], S. S. P. Parkin[1], Y. L. Chen[2,6,7,9†]

[1]*Max Planck Institute of Microstructure Physics, Halle, 06120, Germany*
[2]*School of Physical Science and Technology, ShanghaiTech University, Shanghai, 201210, China*
[3]*Institute of Physics, Chinese Academy of Sciences, Beijing, 100190, China*
[4]*Songshan Lake Materials Laboratory, Dongguan, Guangdong, 523808, China*
[5]*Max Planck Institute for Chemical Physics of Solids, Dresden, D-01187, Germany*
[6]*Clarendon Laboratory, Department of Physics, University of Oxford, Oxford OX1 3PU, U.K.*
[7]*ShanghaiTech Laboratory for Topological Physics, Shanghai 200031, P. R. China*
[8]*Diamond Light Source, Didcot, OX110DE, U.K.*
[9]*State Key Laboratory of Low Dimensional Quantum Physics, Department of Physics, Tsinghua University, Beijing 100084, China*
[10]*Frontier Science Center for Quantum Information, Beijing 100084, China*
[11]*John A. Paulson School of Engineering and Applied Sciences, Harvard University, Cambridge, MA 02138, USA*
[12]*Department of Physics, Harvard University, Cambridge, MA 02138, USA*

\* These authors contributed equally to this work
† Email: yulin.chen@physics.ox.ac.uk

The spin-orbit coupling (SOC) lifts the band degeneracy that plays a vital role in the search for different topological states, such as topological insulators (TIs) and topological semimetals (TSMs). In TSMs, the SOC can partially gap a degenerate nodal line, leading to the formation of Dirac/Weyl semimetals (DSMs/WSMs). However, such SOC-induced gap structure along the nodal line in TSMs has not yet been systematically investigated experimentally. Here, we report a direct observation of such gap structure in a magnetic WSM Co$_3$Sn$_2$S$_2$ using high resolution angle-resolved photoemission spectroscopy. Our results not only reveal the existence and importance of the strong SOC effect in the formation of the WSM phase in Co$_3$Sn$_2$S$_2$, but also provide insights for the understanding of its exotic physical properties.

The spin-orbit coupling (SOC) effect originates from the relativistic interaction between a particle's spin and its orbital motion, which can modify atomic energy levels and split the energy bands in crystalline materials. The SOC effect is important in numerous research fields of physics, such as spintronics [1], ultracold atoms [2], high temperature superconductivity [3,4] and topological materials [5-7]. In solids, it can lift the degeneracy of critical bands and play a vital role in the formation of exotic topological states [5-7]. As an example, in some TSMs, depending on how the SOC effect modifies the topological nodal line, different topological states can be formed: (i) the topological nodal line semimetals (TNLSMs) state can be formed if the nodal line is not gapped [8,9]; (ii) the formation of topological insulators (TIs) if the nodal line is fully gapped [10-12]; (iii) if the nodal line is partially gapped with isolated nodal points, Dirac semimetals (DSMs) or Weyl semimetals (WSMs) can be formed [13-21].

In TIs and TSMs, the strength of SOC can be characterized by the energy gap size between the inverted bands and is dependent to the atomic mass. In compounds with heavy elements (e.g. bismuth-based TIs), the SOC-induced energy gap up to several hundred millielectron volts has been observed [22-25]. On the other hand, although many TSMs have been discovered [13-21,26-29] up to date, the SOC-induced gap structure along the nodal line has not yet been systematically investigated. However, the recently discovered magnetic WSM $Co_3Sn_2S_2$ [18-21] provides an ideal opportunity for such a study.

The WSM $Co_3Sn_2S_2$ has three pairs of Weyl points formed by partially gapped nodal lines due to the SOC effect [20]; which also give rise to giant anomalous Hall effect [18,19]. The scanning tunneling microscopy (STM) measurements observed a large negative flat band magnetism [30] and spin-orbit polaron [31] in $Co_3Sn_2S_2$ again indicate the effect of strong SOC in $Co_3Sn_2S_2$. Interestingly, a recent photoemission study [32] suggests the SOC effect in $Co_3Sn_2S_2$ is negligible and a Weyl loop state is formed.

In this report, we systematically investigate the SOC-induced gap structure along the nodal line in $Co_3Sn_2S_2$ using high resolution angle-resolved photoemission spectroscopy (ARPES), and directly observed large SOC-induced energy gap (up to ~ 55 meV) distribution in the momentum space, which can be well reproduced by our *ab initio* calculations. These results clearly support the WSM nature, instead of the Weyl loop state in $Co_3Sn_2S_2$; which provide a solid electronic structure foundation for understanding many exotic physical properties in $Co_3Sn_2S_2$, such as the large anomalous Hall conductivity (AHC) [18,19], large anomalous Hall angle (AHA) [18] and anomalous Nernst effect (ANE) [33,34].

$Co_3Sn_2S_2$ is crystallized in a trigonal rhombohedral structure and composed of stacked ...-Sn-[S-($Co_3$-Sn)-S]... layers (Fig. 1a). In each layer, Co atoms form a two-dimensional (2D) kagome lattice. Such unique crystal structure guarantees the inversion symmetry, $C_{3z}$ rotation symmetry and three mirror planes in $Co_3Sn_2S_2$. It undergoes a ferromagnetic (FM) transition at ~177 K [18,19]. The *ab-initio* calculations on $Co_3Sn_2S_2$ in its FM state reveal a band inversion occurs near Fermi level ($E_F$) [20]. When SOC is ignored, the band inversion forms a nodal line as illustrated in Fig. 1c and totally 6 nodal lines locating in three mirror planes of the Brillouin zone (BZ) are formed (Fig. 1b). Each nodal line will be partially gapped when SOC is considered, leaving two nodal points in formation of the Weyl points with opposite chiralities as illustrated in Figs. 1b and 1d.

To quantitatively visualize the position of the nodal line and the SOC effect on the nodal line, we first calculated the band dispersions along $\overline{M}'$ - $\overline{\Gamma}$ - $\overline{M}$ direction with and without SOC as shown in Figs. 2c and 2d, respectively. Their momentum positions in the BZ are shown in Fig. 2a. Without SOC effect, the nodal line is characterized by the linear band crossings for all selected dispersions as shown by the black arrows in Fig. 2c. With changing the $k_z$ value, the crossing point originally lies at ~ 50 meV above $E_F$ [Fig. 2c (i, ii)], then moves down to ~ -50 meV below $E_F$ [Fig. 2c (iii, iv)], and eventually moves up to ~ 100 meV above $E_F$ [Fig. 2c (v, vi)]. Such behavior indicates the strong energy dispersion of the nodal line in the mirror plane. When SOC is considered, the crossing points are gapped as shown in Fig. 2d (i, iii-vi), while it remains nodal in Fig. 2d (ii) forming the Weyl point. We illustrate the SOC-induced energy gap size along the nodal line in Fig. 2b. It shows strong anisotropy ranging from 0 to ~ 50 meV. As ARPES probes the occupied states below $E_F$, the SOC-induced gap structure in the Fig. 2d (iii, iv) can be observed experimentally. Based on the calculations, the portion of the gapped nodal line lying below $E_F$ are marked by the cyan region as illustrated in Fig. 2a.

To search for the gapped structure along the nodal line, we carried out detailed photon energy dependent measurements along $\overline{\Gamma}\overline{M}$ direction as shown in Fig. 3. The experimental dispersions taken at different photon energies are plotted in Fig. 3a. The measured position on the nodal line at different photon energies is illustrated by the black line in the inset of each panel. At the photon energy of 120 eV which accesses the $k_z = 0$ [Fig. 3a (i)], the gapped structure lies above $E_F$ that can not be observed. Increasing the photon energy to 125 eV [Fig. 3a (ii)], the measured point on the nodal line lies in the cyan region [see the inset of Fig. 3a

(ii)] indicating the gapped structure lying below $E_F$. Indeed, the corresponding gapped structure is clearly observed experimentally, characterized by the loss of intensity as illustrated by the arrow in Fig. 3a (ii). More photon energy dependent measurements on different positions along the nodal line in the cyan region also resolve the SOC-induced gap structure [Fig. 3a (iii - vi)]. As the measured position moves out of the cyan region [see the inset of Fig. 3a (vii)], the gapped structure lies above $E_F$ again. It is also confirmed by the experimental dispersion taken by 141 eV [Fig. 3a (vii)]. The energy distribution curves (EDCs) corresponding to Fig. 3a are shown in Fig. 3b. The SOC-induced gap is clearly visualized between the two branches of bands [Fig. 3b (ii - vi)].

We also carried out high resolution measurements on the Fermi surface (FS) topology using the photon energy of 125 eV (Fig. 4a) and 134 eV (Fig. 4c) where the gapped nodal line lying below $E_F$. Both FS topologies exhibit a 3-fold symmetry which is consistent with the crystal structure. Three spot-like features were observed along $\overline{\Gamma}\,\overline{M}$ direction on both FSs in the first BZ. These features originate from the upper branch of the gapped nodal line. In addition to the SOC-induced gap observed along the $\overline{\Gamma}\,\overline{M}$ direction (Fig. 3), we extracted the band dispersion through the spot-like feature perpendicular to the $\overline{\Gamma}\,\overline{M}$ direction as shown in Figs. 4b and 4d. Apparently, an energy gap developed between the upper and lower branch of the bands is clearly observed as illustrated by the arrows in Figs. 4b (i) and 4d (i). Such gap structure can also be seen from the EDCs, characterized by the two peaks structure with a dip in the middle [Figs. 4b (ii) and 4d (ii)].

By tracking the band dispersions, the position of the nodal line were determined experimentally as shown in Fig. 4e. It shows excellent agreement with the calculations. To quantitatively extract the SOC-induced gap size along the nodal line, we fit the two peaks of the EDCs by using two Lorentzian curves [see the insets of Figs. 4b (ii) and 4d (ii) for example]. The gap size is extracted by the energy interval between the two peaks and the results are shown in Fig. 4f. The gap size measured in the cyan region is around 55 meV, which is consistent with the calculations (Fig. 2b).

The direct observation of the SOC-induced gap structure along the nodal line in $Co_3Sn_2S_2$, together with the excellent agreement with *ab initio* calculations unambiguously demonstrate the WSM nature of $Co_3Sn_2S_2$ with isolated Weyl points, instead of the degenerate Weyl loops. These results not only reveal the existence and importance of the strong SOC effect in the formation of the WSM phase in $Co_3Sn_2S_2$, but also provide important insights for the understanding of many exotic physical properties in $Co_3Sn_2S_2$, such as the large AHC [18,19], AHA [18] and ANE [33,34].

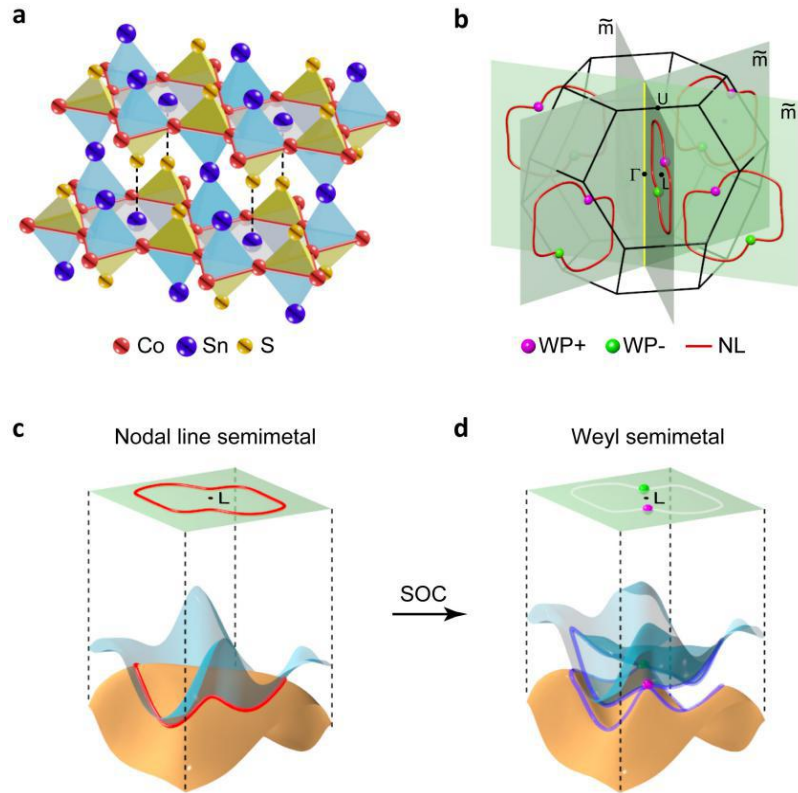

**Fig. 1. Magnetic Weyl semimetal phase in Co$_3$Sn$_2$S$_2$.** (a) The crystal structure of Co$_3$Sn$_2$S$_2$. (b) Schematic of the nodal line and Weyl points lying in three mirror planes of the bulk Brillouin zone. Red line represent the nodal line (NL) in the absence of the spin-orbit coupling (SOC) effect. In the presence of SOC, the nodal line will be partially gapped, leaving two nodal points forming the Weyl points (WPs) with positive (+, the magenta color point) and negative (-, the green color point) chirality, respectively. (c) Illustration of the nodal line originating from the band inversion in the absence of the SOC. (d) Illustration for the formation of the Weyl points and the partially gapped nodal line when SOC is included.

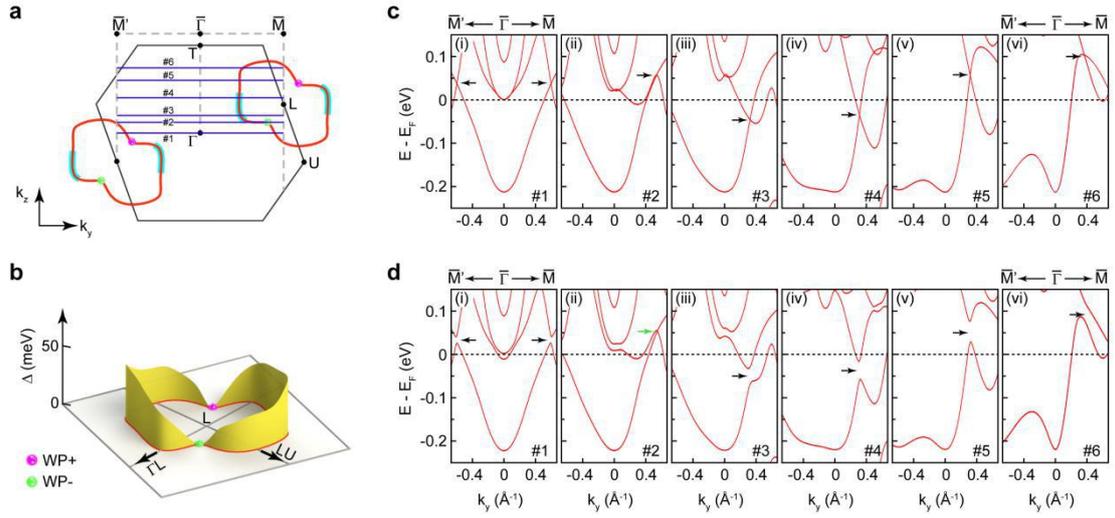

**Fig. 2. The calculated band dispersions without and with SOC.** (a) Illustration of the nodal line and Weyl points in one mirror plane. The cyan region of the nodal line represents the SOC-induced gap structure lying below $E_F$. (b) The calculated SOC-induced gap size along the nodal line. The gap size is renormalized by a factor of 1.43. (c) The calculated band dispersions through the nodal line without SOC. The calculated bandwidth was renormalized by a factor of 1.43 and the energy position was shifted to match the experiments. Their momentum positions are shown by the blue lines in (a). The black arrows illustrate the nodal points along the nodal line. (d) The corresponding calculated band dispersions of (c) with SOC. The black arrows illustrate the gap opened by SOC effect. The green arrow in (ii) illustrates the Weyl point.

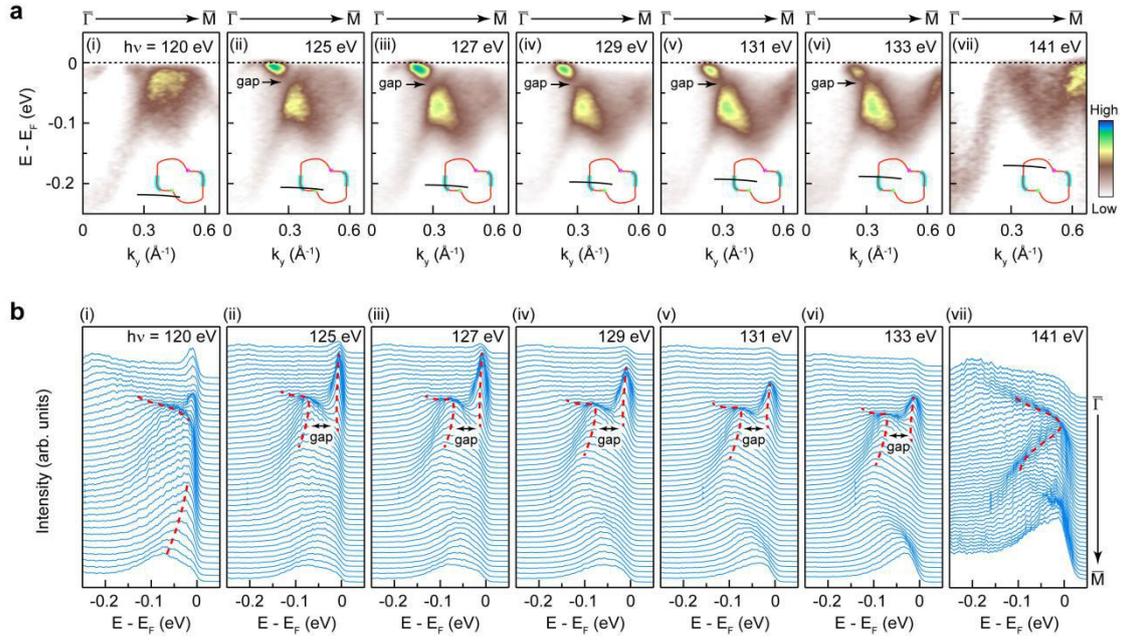

**Fig. 3. Observation of the SOC-induced gap structure along the nodal line in $Co_3Sn_2S_2$.** (a) The band structure along $\overline{\Gamma}\,\overline{M}$ direction taken at different photon energies. The measured position on the nodal line is marked by the black line in the inset of each panel. When the measured points locate out of the cyan region on the nodal line (i, vii), the gapped nodal line lies above $E_F$ that can not be observed experimentally. Upon moving the measured point into the cyan region, the gapped nodal line lies below $E_F$, that are clearly observed experimentally as illustrated by the arrows in (ii - vi). (b) The corresponding energy distribution curves (EDCs) of (a). The SOC-induced gap is characterized by the separation of the two branches of bands in (ii - vi).

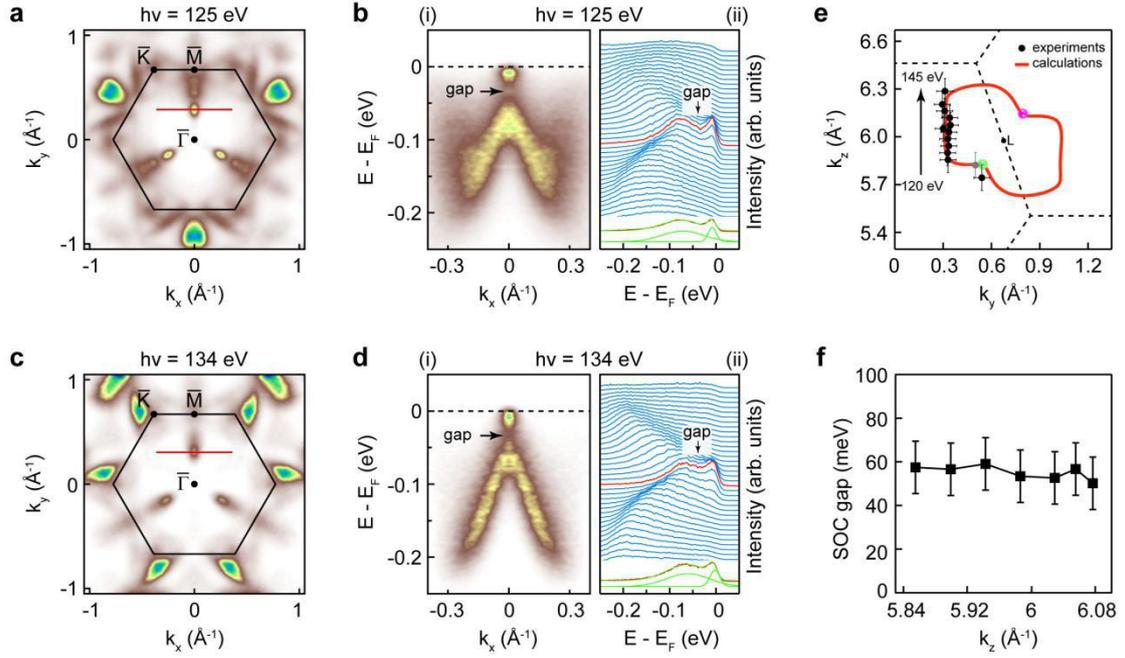

**Fig. 4. SOC-induced gap size along the nodal line.** (a) The Fermi surface topology taken at 125 eV photon energy. Three spot-like features are observed along $\bar{\Gamma}\bar{M}$ direction which originate from the upper branch of the gapped nodal line. (b) The extracted band structure (i) and its corresponding EDCs (ii) perpendicular to the $\bar{\Gamma}\bar{M}$ direction through the spot-like feature. The momentum path is illustrated by the red line in (a). The SOC-induced gap is clearly observed as illustrated by the arrow. To quantitatively extract the gap size, the EDC in red is fitted by using two Lorentzian curves as illustrated by the green curves in the inset of (ii). (c, d) The same to (a, b) but the data is taken at 134 eV photon energy. (e) The extracted positions of nodal line. The gray point is taken by 115 eV photon energy and the position is obtained based on the inversion symmetrization. The experimental results show excellent agreement with the calculations. (f) The experimentally extracted SOC-induced gap size along the nodal line as a function of $k_z$.